\documentstyle[a4,11pt]{article}
\newcommand \beq{\begin{eqnarray}}
\newcommand \eeq{\end{eqnarray}}
\def\leftrightarrowfill{$ \mathord\leftarrow \mkern-6mu \cleaders
\hbox{$\mkern-2mu \mathord- \mkern-2mu$}\hfill \mkern-6mu \mathord\rightarrow$}
\def\overleftrightarrow#1{ \vbox{\ialign{##\crcr \leftrightarrowfill\crcr
\noalign{\kern-1pt\nointerlineskip}
$\hfil\displaystyle{#1}\hfil$\crcr}}}
\def\sqr#1#2{{\vcenter{\vbox{\hrule height.#2pt \hbox {\vrule width.#2pt
height#1pt \kern#1pt \vrule width.#2pt} \hrule height.#2pt }}}}

\begin{document}
\title{\sf Enhancement of nucleon-nucleon cross section in sidewards flowing
nuclear matter through zero sound mode excitation}
\author{\sf J. Diaz-Alonso$^*$, L. Mornas$^\dagger$ \\ 
\\
{\small{\it $^*$DARC, UPR 176 CNRS, Observatoire de Paris - Meudon, 
F-92195 Meudon, France}} \\
{\small{\it and Universidad de Oviedo, Departamento de F\'{\i}sica,
E-33007 Oviedo, Spain}} \\
{\small{\it $^\dagger$Departamento de F\'{\i}sica Te\'orica, 
Universidad de Valencia, 46100 Burjasot (Valencia), Spain }} }
\maketitle
\par\noindent {\small keywords: in-medium NN cross section, screened one-boson 
          exchange,relativistic RPA, zero-sound}

\begin{abstract} 
We investigate the modification induced on the nucleon-nucleon 
cross section by screening of the interaction inside nuclear matter.
The interaction is described by the relativistic one boson exchange 
of $\sigma$, $\omega$ and $\pi$ mesons. Medium effects enter as loop 
corrections to the meson propagators and are characterized not only by 
density, but also by temperature and velocity with respect to the center 
of mass of the collision. The cross section displays peaks for some 
specific values of the velocity, corresponding to the excitation of 
zero-sound modes of the longitudinal $\omega$ field. The enhancement 
factor amounts to about 1.5 under reasonable conditions. It increases 
with density and is reduced at finite temperature. These findings may 
have verifiable consequences on the angular dependence of the measurable 
secondary particle distributions. 
\end{abstract}


\section{Introduction}
\label{intro}
Dissipative effects play an important role in explaining the shapes of the 
angular distributions and energy spectra of the secondary particles produced 
in a relativistic heavy ion collision. For this reason, the cross section is 
an important ingredient in all kinds of numerical simulations of these 
collisions, be it in the molecular dynamics approach \cite{RQMD,AMD}, kinetic 
theory approaches \cite{BUU,KWB93}, or the hydrodynamical models where it 
enters through the transport coefficients \cite{hydro}. It has become 
obvious in the last years that using the free value of the cross section 
does not well reproduce the available experimental data; rather, the medium 
may affect the nucleon-nucleon cross section by modifying the strength of 
the interaction and the available phase space.

Several attempts to calculate medium corrections to the cross section
already exist. The first point to precise is in what type of background
matter does the collision take place. Some authors like Faessler
{\it et al.} \cite{Faessler} or Fuchs {\it et al.} \cite{Fuchs}
consider that the background consists of a momentum distribution 
in two well separated Fermi spheres. This corresponds to the initial 
stage of a heavy ion collision and is far out of equilibrium. 
Other authors like Machleidt {\it et al.}  \cite{Machleidt} 
or Alm {\it et al.} \cite{Alm} consider that the background 
is nuclear matter, supposedly of infinite extension, and not far 
from equilibrium. This rather corresponds to the intermediate 
stage of the collision after a fireball was formed. This latter case 
is what we are interested in here.

Medium effects enter at several levels of the calculation of the
cross section. Older evaluations only took into account 
the shifting of the nuclear mass to an effective smaller 
value inside nuclear matter. Then, in the last few years were
performed a number of calculations with the Brueckner formalism
\cite{Machleidt,Alm,Malfliet}. In this approach, an infinite
series of ladder diagrams are resummed. The nucleon propagators,
{\it i.e.} the uprights of the ladder, are dressed. The medium
effects enter as Pauli blocking factors on the intermediate
states in the inner nucleon lines. The meson lines are the free
ones, corresponding to an unscreened interaction.

On the other hand, we would like to discuss in this letter the influence 
of the screening of the interaction by the surrounding nuclear matter. 
In a model where the N-N interaction is described by 
relativistic meson exchanges, we take into account the dressing of the 
meson lines in the first Born approximation. A detailed study of 
the density and temperature dependence of the total and differential 
elastic $p$-$p$ cross section was performed in our first paper 
\cite{ourpaper}. It was assumed there that the background nuclear matter 
was at rest with respect to the center of mass of the collision. The
concern of our present work is to release this assumption. We refer the 
reader to \cite{ourpaper} for details on the formalism and only recall 
here the main formulaes. 


\section{The model}
\label{model}

We are working with the standard quantum hadrodynamics lagrangian 
with $\sigma$, $\omega$ and $\pi$ meson exchange. This is the minimal model 
which permits to reproduce the available experimental data on $p$-$p$ 
scattering to a satisfactory level of accuracy. The interaction lagrangian 
reads
\beq
{\cal L}\, = \, g_{\sigma}\,\overline{\psi}\,{\sigma}\,\psi\,
    +\,g_{\omega}\,\overline{\psi}\,\gamma^{\mu}\,{\omega}_{\mu}\,\psi\,
    +\,g_{\pi}\,\overline{\psi}\,\gamma^{5}\,\vec{\tau}\vec{\pi}\,\psi  
\label{eq:(1)}
\eeq
The differential elastic cross section was calculated in the Born 
approximation. In the center of mass of the collision,
\beq
   d \sigma / d \Omega = 1/(64 \pi^2 s) |{\cal M}|^2
   \label{eq:(2)}
\eeq
where $s$ is the Mandelstam variable $s=4 E_{c.m.}^2$ and the transition
matrix is given by
\beq
   {\cal M}=\sum_{{\rm m,n}=\sigma,\omega,\pi} (\overline U_3 \Gamma_m U_1) 
   G_{31}^{\rm m n} 
   (\overline U_4 \Gamma_n U_2) - \hbox{\rm exchange} 
   \label{eq:(3)}
\eeq
In this expression the $U$'s are the spinors defining the incoming 
($ U_1, U_2 $) and outgoing ($ U_3, U_4 $) nucleon states. The indices
$i=1,2,3,4$ carried by the spinors stand for the momenta $p_i^\mu$ and 
the spin-isospin quantum numbers $s_i$, $t_i$. $G_{31}^{\rm mn}$ is the 
propagator matrix  which depends on the momentum transfer $k_{31}^\mu=
p_1^{\mu}-p_3^{\mu}$. In free space, it reduces to the standard form 
\beq
G_{31}^{\rm mn}=\mbox{\rm Diag}\left( \ {-i [ g^{\mu\nu} - {k^\mu k^\nu \over 
k^2} ] \over k^2-m_{\omega}^2},\ {i \over k^2-m_{\sigma}^2},\ 
{i \over k^2-m_{\pi}^2}\ \right)
\label{eq:(4)}
\eeq
$ \Gamma $ is the coupling matrix 
$ \Gamma = \left( -i g_{\omega} \gamma^{\mu},\  i g_{\sigma} ,\
                      g_{\pi} \gamma_5 \vec \tau
   \right)$.
In the actual calculation of the transition matrix, all coupling constants 
are multiplied by form factors $(\Lambda_{\rm m}^2 - 
m_{\rm m}^2)/(\Lambda_{\rm m}^2 -  k^2)$. The exchange term is obtained by 
interverting indices 3 and 4. In the following we consider only the 
spin-averaged cross section 
$\overline{|{\cal M}|^2}=(1/4) \sum_{s_i} |{\cal M}|^2$.  
$\sigma_{pp}$ and $\sigma_{nn}$ are equal in our approximation where the 
Coulomb interaction was substracted and the neutron and proton masses are 
equal. We do not show $\sigma_{np}$ since our simple model does not include 
$\rho$ meson exchange. 
\par
In paper I we performed a fit of the free elastic $p$-$p$ cross section 
using the available experimental data \cite{data}. We kept constant 
the masses of the mesons and the pion-nucleon coupling ($m_{\sigma}$=550 MeV, 
$m_{\omega}$=783 MeV, $m_{\pi}$=550 MeV, $g_{\pi}^2/4 \pi$=14.4) and 
fitted the coupling constants $g_{\sigma}$, $g_{\omega}$ and the cutoffs 
$\Lambda_{\sigma}$, $\Lambda_{\omega}$, $\Lambda_{\pi}$.  Several sets 
of parameters were found to reproduce the data to a satisfactory level of 
accuracy. It was checked that the in-medium behavior of the cross section
was the same for all choices of the parameter set. The calculations are 
performed here with parameter set A of paper I: $g_{\sigma}=3.80$, 
$g_{\omega}=9.31$, $\Lambda_{\sigma}=1298.8$, $\Lambda_{\omega}=1240.5$, 
$\Lambda_{\pi}=362.1$.

\par
Inside the medium, the momenta and masses of the incoming and outgoing
nucleons must be replaced by their effective counterparts. In the
Hartree approximation, $p_i^\mu \rightarrow (p_i^\mu)^*=p_i^\mu-g_\omega$ 
$<\! \omega^\mu \! >$, $m \rightarrow m_*=m -g_\sigma <\! \sigma \! >$. 
The propagator matrix must also be modified in order to take into account 
that the N-N interaction is mediated by dressed mesons.
The dispersion relations and propagators of the mesons at finite temperature 
were obtained from a linear kinetic analysis of perturbations around the
Hartree ground state \cite{DP91}. 
The results of this method coincide at 
$T$=0 with those of the one-loop approximation \cite{GDP94}. The calculation 
takes into account renormalized vacuum polarization contributions which are 
crucial in obtaining a physically reasonable behavior of the propagation 
modes \cite{DP91}.
The propagator matrix so obtained has the form
\cite{ourpaper,GDP94,DFH92}.
\beq
   G^{mn}=\left(
   \begin{array}{ccc} G_{\omega}^{\mu\nu} & G_{\sigma\omega}^{\mu} & 0 \\ 
                      G_{\omega\sigma}^{\mu} & G_{\sigma} & 0 \\
                      0 & 0 & G_{\pi}  
   \end{array}\right)
   \label{eq:(5)} 
\eeq
\beq
{\rm with}\qquad  
G_{\omega}^{\mu\nu} & = & -i \Bigl[ {\cal T}^{\mu\nu}\ {1 \over 
     k^2 -m_{\omega}^2 +g_\omega^2 \Pi_{\omega T}} 
   + \Lambda^{\mu\nu}\ {k^2 -m_{\sigma}^2 + \ g_{\sigma}^2 \Pi_{\sigma}
     \over D_{\sigma-\omega L}(k)} 
   - {k^{\mu} k^{\nu} \over m_\omega^2 k^2}  \Bigr]  \nonumber \\
G_{\sigma} & =& i {k^2-m_{\omega}^2+g_\omega^2 \Pi_{\omega L}
   \over D_{\sigma-\omega L}(k)}  \quad ; \quad
G_{\sigma\omega}^{\mu}  =  -i \eta^{\mu} {g_\sigma g_\omega \ 
    \Pi_{\times} \over D_{\sigma-\omega L}(k)}  \nonumber \\
G_{\pi}& =& i {1 \over k^2 -m_{\pi}^2 + \ g_{\pi}^2 \Pi_{\pi}} 
   \label{eq:(6)}
\eeq
In these expressions we defined the quadrivector $\eta^\mu$ and 
the projectors onto the longitudinal ($\Lambda^{\mu\nu}$) and transverse 
(${\cal T}^{\mu\nu}$) modes as functions of the 4-velocity $u^\mu$ of the 
fluid as follows:
\beq
\eta^{\mu} & = & u^{\mu} - {k.u \over k^2} k^{\mu} \qquad ; \qquad
\tilde g^{\mu\nu}  =  g^{\mu\nu} - {k^\mu k^\nu \over k^2} \nonumber \\
{\cal T}^{\mu\nu} & = & \tilde g^{\mu\nu} - {\eta^{\mu} \eta^{\nu} \over 
\eta^2} \qquad ; \qquad \Lambda^{\mu\nu} ={\eta^\mu\eta^\nu \over \eta^2}
   \label{eq:(7)}
\eeq
The polarizations $\Pi(k)$ of the meson fields are functions of the 
thermodynamical state (density and temperature) of the matter
and on the transferred 4-momentum $k^\mu=(\omega,\vec q)$. There is a 
mixing between the $\sigma$ and longitudinal-$\omega$ propagation modes. 
The $\pi$ propagation is decoupled from the dynamics of the other mesons 
in symmetric nuclear matter. We use the notation 
\beq
\Pi_{\times} &=& \eta_{\mu} \Pi_{\sigma\omega}^\mu \qquad ; \qquad
\Pi_{\omega L}= - \Lambda_{\mu\nu} \Pi_\omega^{\mu\nu} \qquad ; \qquad
\Pi_{\omega T}= - {\cal T}_{\mu\nu} \Pi_\omega^{\mu\nu} /2
  \label{eq:(8)}
\eeq
for the mixing, longitudinal and transverse parts. 
The explicit expression 
for the polarizations and a study of the dispersion relations can found 
in \cite{ourpaper,DP91}. Besides the $\sigma$ and longitudinal and
transverse $\omega$ branches, there is a zero sound branch in the
mixed $\sigma$-$\omega$ mode. 
\par
The denominator of the $\sigma$, longitudinal part of the $\omega$
and mixing propagators is given by the mixed $\sigma$-$\omega$
dispersion relation
\beq
D_{\sigma-\omega L}(k)=
(k^2 -m_{\sigma}^2 + g_{\sigma}^2 \Pi_{\sigma})(k^2 -m_{\omega}^2 
   +g_\omega^2 \Pi_{\omega L} ) +g_\sigma g_\omega \Pi_{\times}^2
   \label{eq:(9)}
\eeq
\par The polarizations are complex. In the $ T = 0 $ limit, the 
imaginary part of the polarizations vanish 
outside the region defined by
\beq
     \sqrt{(p_F-q)^2  + m_*^2} - \sqrt{p_F^2 +  m_*^2} < \omega <
     \sqrt{(p_F +  q)^2 + m_*^2} - \sqrt{p_F^2 + m_*^2}
     \label{eq:(10)}
\eeq
(where $p_F$ is the Fermi momentum and $m_*$ the effective mass of the
nucleon inside the medium). Inside this region, the propagation modes 
are damped by the decay into particle-hole pairs. At finite
temperature the imaginary parts are finite for any space-like mode.
\par
The dependence on the velocity of the medium comes through the quadrivector
$\eta^{\mu}$. In the referential (R1) where the fluid is at rest, 
$u^{\mu} =(1,\vec 0)$ and the momentum transfer is $k^{\mu} = (\omega,\vec q)$

In the center of mass of the collision (referential (R2)), the background 
fluid moves with a velocity $u^{\mu}=(\gamma, \gamma \vec v)$
parametrized by
\beq
\vec v = \left( v \sin \alpha \sin \varphi , 
                        v \sin \alpha \cos \varphi , 
                        v \cos \alpha  \right)
   \label{eq:(11)}
\eeq
The momentum transfers for the direct ($k_{31}$) and exchange ($k_{41}$) 
diagrams are
\beq
k_{31}^{\mu} &=& (0,0,p \sin \theta,p (1-\cos\theta)) \qquad ; \qquad
k_{41}^{\mu} =(0,0,- p \sin \theta,p (1+\cos\theta)) 
   \label{eq:(12)}
\eeq
with $p=\sqrt{E_{c.m}^2+m_*^2}$ and $\theta$ is the $c.m.$ 
scattering angle.

One goes from (R1) to (R2) by performing a boost of four-velocity
$(\gamma, \gamma \vec v)$ and a rotation.
In the referential (R1), 
\beq
k_{31}^\mu (R1) & = &  (\omega_\ominus,\vec q_\ominus) \qquad ; \qquad
k_{41}^\mu (R1) =  (\omega_\oplus,\vec q_\oplus) \nonumber \\
\omega_{({\ominus \atop \oplus})} & = & \gamma v p [\pm \sin \alpha 
\cos \varphi \sin \theta - \cos \alpha (1 \mp \cos \theta)]  \\
   \label{eq:(13)}
q_{({\ominus \atop \oplus})} & = & [\omega_{({\ominus \atop \oplus})}^2 +
2 p^2 (1 \mp \cos \theta) ]^{1/2} \nonumber 
\eeq


\section{In medium cross section for finite relative velocity}
\label{finitevel}

When the rest frame of the background nuclear matter is moving
with respect to the center of mass of the colliding particles, 
the relative velocity will enter the expression of the cross section 
through the meson propagators. There is an explicit dependence coming 
from the projectors $\Lambda^{\mu\nu}$, ${\cal T}^{\mu\nu}$ (see 
Eq.~(\ref{eq:(7)})). The projectors act by attributing velocity-dependent 
weights to the various components of the currents $\dot{\cal J}$ like 
{\it e.g.} $\dot{\cal J}_\omega^{\mu\nu}=(\overline U_3 \gamma^\mu U_1)\, 
(\overline U_4 \gamma^\nu U_2)$. There is also an indirect dependence in 
the velocity in the expression of the polarizations 
$\Pi(\omega_{({\ominus \atop \oplus})},q_{({\ominus \atop \oplus})})$ 
coming from the boost transformation Eq.~(13). In contrast
to the zero velocity case where the momentum transfer $k^\mu$ entering
the expression of the polarizations was confined to the axis $\omega$=0,
$k^{\mu}=(\omega_{({\ominus \atop \oplus})},q_{({\ominus \atop \oplus})})$  
now spans all the spacelike region. We will consider three cases:

\medskip
\par\noindent {\it (i)} $\underline{v=(0,0,v_z)}$ \par\noindent
For moderate values of the velocity along the axis defined by the 
incident particles ($v=v_z$), the cross section is slightly 
enhanced at low densities and reduced at high densities. 
When some temperature is introduced, the low density enhancement
disappears and the high density reduction persists. At highly 
relativistic $v_z$ velocities ($\gamma_z \stackrel{>}{\sim} 5$), 
the cross section displays a series of peaks, of which three are 
clearly visible. The height of the peaks increases with density.
The effect of finite temperature is to gradually wash out these 
peaks. The peaks are superposed on an additional increase of
the cross section, which stabilizes to a large constant value
for large $\gamma_z$. This additional enhancement persists
when introducing finite temperature.

\medskip
\par\noindent {\it (ii)} $\underline{v=(0,v_y,0)}$ \par\noindent
For low values of the velocity inside the reaction plane and 
orthogonal to the beam axis ($v=v_y$), the cross section is
first approximately constant for densities up to $n=n_0$ and 
somewhat reduced at high density when increasing $v_y$.
Further increasing $v_y$, one meets a peak structure (see
Fig.~1) superposed on a smooth increase similar to the one 
already observed in the case $v=v_z$ , but occurring for more 
physical values of the velocity. There is first a  sharp 
peak for a characterictic value $\hat v_y$ immediately followed 
by one or two smoother peaks. Further increasing the velocity, 
the cross section slowly decreases and stabilizes to a constant 
value $\sigma_{med}/\sigma_{free} (v_y \rightarrow c)$
somewhat larger than $\sigma_{med}/\sigma_{free} (v=0)$ and
approximately equal to 1. We were able to check that the peak
structure originates from the velocity dependence of the 
polarizations while the residual temperature-independent 
enhancement at $v_y \rightarrow c$  is due to the velocity 
dependence of the projectors $\Lambda^{\mu\nu}$, ${\cal T}^{\mu\nu}$. 
The same holds for the case $v /\! / z$.

\medskip
\par\noindent {\it (iii)} $\underline{v=(v_x,0,0)}$ \par\noindent
No appreciable modification is observed for velocities orthogonal 
to the reaction plane ($v=v_x$) up to $v_x \simeq 0.8 c$. On the other
hand, the total cross section starts to increase above this value 
and continues to grow indefinitely as $\gamma_x \rightarrow \infty$.

The angular dependence of the differential cross section deviates 
appreciably from its behavior when the fluid is at rest. 
For each value of the velocity, there is a favoured angle,
but this effect is most appreciable for small angles,
so that finite velocities also contribute to make the cross section
more forward/backward peaked with respect to its vacuum value.

We concentrate on the behaviour with $v_y$ which is the case most liable 
to lead to observable consequences since the peaks appear for reasonable 
values of the relative velocity. 
The characteristic velocity $\hat v_y$ at which the sharp peak takes 
place is density dependent, {\it e.g.} one has $\hat v_y \simeq 0.41 c$ 
for $n$=$n_0$ and $\hat v_y \simeq 0.57 c$ for $n$=$2 n_0$ (see Fig.~1a). 
The height of the peak increases with increasing density and is 
reduced by increasing the temperature (see Fig.~1b). At $T$=10 MeV, 
a broad bump is still clearly visible around $\hat v_y$, at high 
temperature $T \stackrel{>}{\sim}$ 50 -- 100 MeV it has completely 
disappeared.

The peak structure is to be related to the excitation of the zero sound 
modes. These
modes arise from the mixed $\sigma$ - longitudinal $\omega$ part of the
dispersion relation; they were studied in previous work \cite{DP91,DFH92}. 
A contour plot of the mixed dispersion relation $D_{\sigma-\omega L}
(\omega,q)$ in the $q$ -- $\omega$ plane of transferred momentum is shown 
in Fig.~2 for $T$=0, $n$=$n_0$. The level $D_{\sigma-\omega L}(\omega,q)=0$ 
gives the location of the zero-sound branch.
For not too high values of the energy-momentum transfer, the region of 
the zero sound branches can be crossed for some values of the velocity
in the medium: If $(\omega,q)$ coincides with a zero-sound solution 
$k^\mu_{zs}=(\omega_{zs},q_{zs})$ of the dispersion relation 
$D_{\sigma-\omega L}(k^\mu_{zs})=0$, there will be 
a critical value of the relative velocity $\hat v$ solution of 
\beq
\omega_{zs}=\omega_{({\ominus \atop \oplus})} = \gamma \hat v p 
[\pm \sin \alpha \cos \varphi \sin \theta - \cos \alpha (1 \mp \cos \theta)]
\eeq
({\it cf.} Eqs.~(\ref{eq:(9)},13)) for which the real part of the dispersion 
relation vanishes. Since the dispersion relation enters in the denominator of 
the propagators ({\it cf.} Eq.~(\ref{eq:(7)})), a pole appears in the 
expression of the cross section ({\it cf.} Eqs.~(\ref{eq:(2)},\ref{eq:(3)})) 
if the imaginary part of $D_{\sigma-\omega L}$ 
vanishes as well. For velocities near $\hat v$, 
$D_{\sigma-\omega L}$ is still reduced and as a consequence, the 
cross section enhanced.

For example, at $n$=$n_{0}$ and $T$=0, we have drawn on Fig.~2 the path 
swept by the components of momentum transfer for fixed laboratory energy 
($T_{lab}$=300 MeV) and relative velocity as the integration needed 
to calculate the total elastic in-medium cross section is carried out. 
It is seen that it falls on the upper part of the zero sound branch for 
$\hat v_y$=0.41 corresponding to the location of the peak in Fig.~1.

The lower branch of the zero sound falls in the particle-hole damping 
zone. On the other hand, part of the upper branch falls outside 
the particle-hole damping zone where the imaginary part of 
$D_{\sigma-\omega L}$ vanishes corresponding to undamped modes at 
our level of approximation. This is at the origin of the narrowest 
peaks shown in Fig.~1, while the smoother peak is to be related to 
the lower branch for which particle/hole damping is active. The 
zero-sound branches become more important as the density increases 
above a threshold value, their importance is reduced as temperature 
increases. It is indeed to be seen in Fig.~1a that the peak structure 
begins to develop above $n$=$0.5\, n_0$ and in Fig.~1b that it is washed 
out at high temperature.


\section{Discussion and conclusion}
\label{conclusion}

In a relativistic OBE model of the N-N interaction with $\sigma$, 
$\omega$ and $\pi$ meson exchange, we investigated the observable 
consequences of a modification of the elastic proton-proton cross 
section inside symmetric nuclear matter by screening of the interaction, 
{\it i.e.} by approximating the meson propagators at the RPA level.
In particular we concentrated on the influence of a non vanishing relative 
velocity between the center of mass of the collision and the rest frame of 
the fluid. The velocity dependence was found to be non monotonous. For 
moderate values of the velocity, 
the cross section is enhanced at low densities and reduced at high densities 
if the velocity is inside the reaction plane $(y,z)$ and no appreciable 
modification is observed for velocities orthogonal to the reaction plane 
($v /\! / x$). 

The main result of our calculations is the enhancement introduced 
by the excitation of the zero sound modes of the medium on both 
the differential and the total elastic cross sections. A peak appears
for a charateristic value of the relative velocity. Its location is 
density dependent; the peak is stronger at higher density and gradually 
blurs out when increasing temperature. It occurs for $v / \! / y$ at 
physically relevant values $0.3 < \hat v_y < 0.8$.
but not for $v / \! / x$ or for unrealistic values of $v / \! / z$. 

Such an asymmetric configuration has not yet been considered up to
our knowledge. A selective enhancement of the cross section for a well 
defined value of the density and velocity will probably have a characteristic 
signature in terms of the transverse flow and angular distributions. 
In a naive hydrodynamical picture, we found that the collisions are more 
efficient when they take place in a background matter flowing in the direction 
corresponding to the ``sidewards flow'', moreover the scattered products
are predominantly directed  backward and forward. However, it is delicate 
at this stage of our investigations to put forward a definite prediction 
on the consequences on experimental observables, since they are the result 
of the interplay of many factors. For example, it is known that viscous 
effects \cite{hydro} or the softening of the equation of state (EOS) through 
a possible phase transition to the quark gluon plasma \cite{eossoft} may 
reduce the directed sidewards flow. We think that our findings may bring 
new arguments in the context of the anisotropic flow observed recently both 
in Au+Au collisions at 100-800 MeV/c at the FOPI facility at GSI Darmstadt 
and by the E877 collaboration at AGS in Au+Au collisions at 10.8 GeV/c. 
The anisotropy is not yet fully understood \cite{anisotrexp} and has been 
analysed mainly in terms of the EOS. It could be worthwhile to investigate 
the effect of an anisotropically enhanced cross section in this kind of 
numerical simulations.

The enhancement we have found only depends on the reduction of the real part
of the dispersion relation associated with the vicinity to a zero sound 
mode in the matter. The existence of a zero-sound branch in nuclear matter 
is well documented (see {\it e.g.} \cite{DP91,zerosound}), so that our result 
is a quite general one. Its size depends on the parameters, in particular 
on the cutoffs $\Lambda_{\sigma}$ and $\Lambda_{\omega}$. Our parameters 
come from a fit of the free cross section; the cutoffs $\Lambda \simeq 1200$ 
MeV and coupling constants are rather lower those of the Bonn potential, so 
that we feel we are on the safe side. 
Let us say a few words on higher order corrections. Vertex functions such as 
calculated by Allendes and Serot \cite{vertex} are amenable at low and 
intermediate momentum transfer to a monopole form factor with $\Lambda 
\simeq 1100$ MeV. If the coupling constants are decreasing functions of the 
density as argued from Dirac-Brueckner-Hartree-Fock calculations, the density 
dependence of our enhancement can be modified. (The determination of a more 
accurate density dependence would require a fully consistent calculation of 
ladders as well as loops which is not yet available.) Even if these branches 
would disappear completely (by applying very low cutoffs), there would still 
be a strong reduction of the dispersion relation $D_{\sigma-\omega L} 
(\omega,q)$ in this zone due to $\sigma-\omega$ mixing and consequently an 
enhancement in the cross section. We believe that this broad enhancement at 
finite relative velocity matters more than the presence of a high but narrow 
peak. Its location will still be determined by the upper limit of the particle
hole damping zone.
 
Finally, we would like to warn the reader against a possible confusion with an
enhancement found by Alm {\it el al.} \cite{Alm} in an other 
context. These authors perform a nonrelativistic Brueckner calculation. 
Their enhancement is caused by a pole in the propagator of the intermediate 
two-nucleon state; the result is interpreted as a precursor of superfluidity. 
The enhancement occurs for low values of the density $n/n_0 \simeq 0.5$; 
the peak is reduced and finally disappears when the total momentum $K$ of 
the pair deviates from zero. This is clearly different from our case where 
the enhancement is due to a pole in the mixed meson propagator and occurs
for high values of the density equal or larger than $n/n_0=0.5$
and for nonvanishing values of the velocity $v/c$= 0.3 - 0.7.

\medskip
\par\noindent{\large{{\bf Figure captions}}}

{\bf Fig.~1}: Ratio of the in-medium cross section with respect
to its value in the vacuum $\sigma_{med}/\sigma_{free}$ as a function 
of velocity (0,$v_y$,0). --- ({\it a}) for density $n/n_0$ = 0.1,
0.5, 1, 2, 5 ---  ({\it b}) temperature dependence at $n=n_0$.

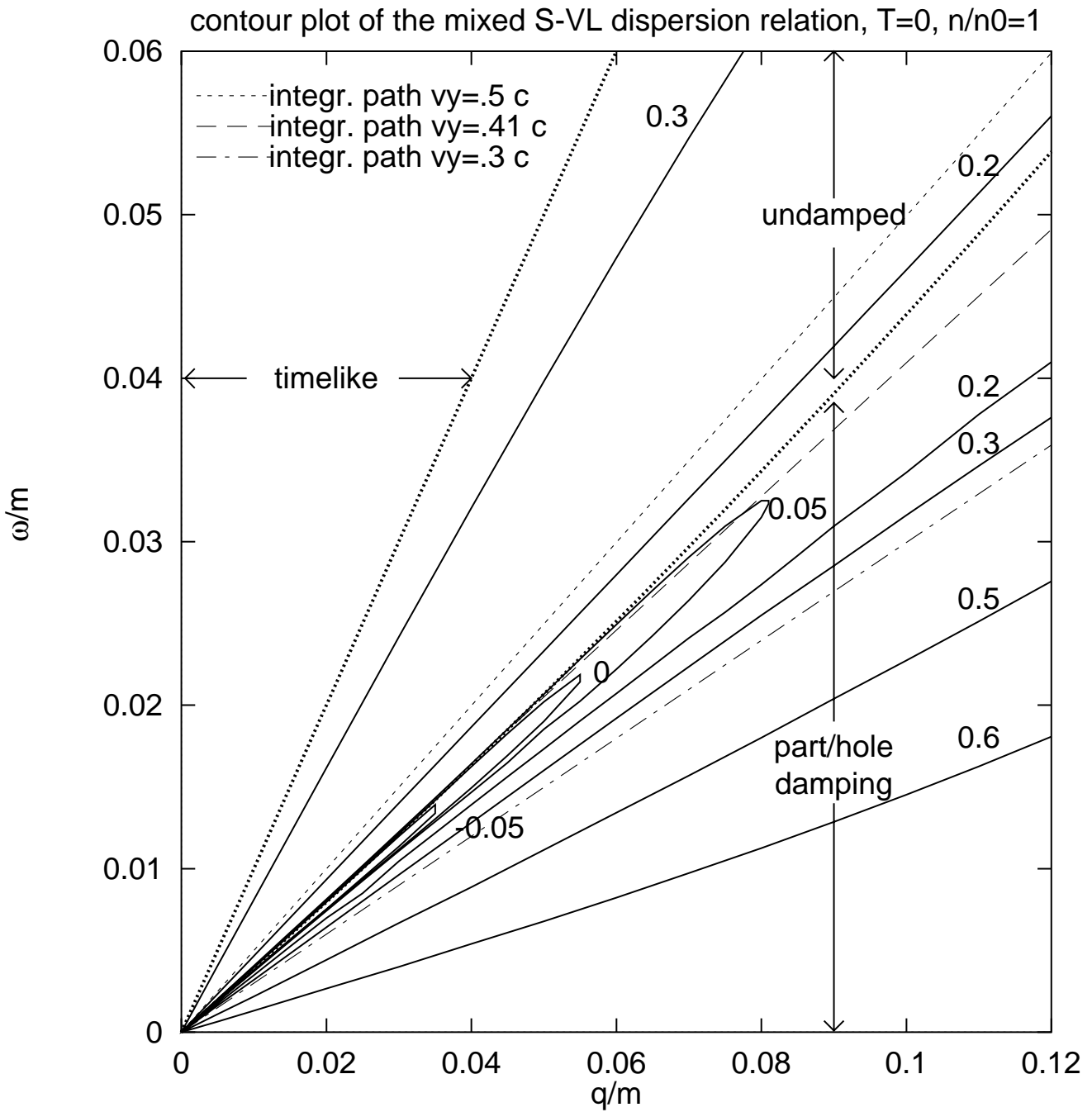
{\bf Fig.~2}: Levels of constant value of the mixed $\sigma$--longitudinal
$\omega$ dispersion relation $D_{\sigma - \omega L}(\omega,q)$ in the 
$q$ -- $\omega$ plane of transferred momentum. The long-dashed line 
$\hat v_y=0.41$ corresponds to the resonant condition $\hat \gamma_y 
\hat v_y p \sin \theta =\omega_{zs}$. 

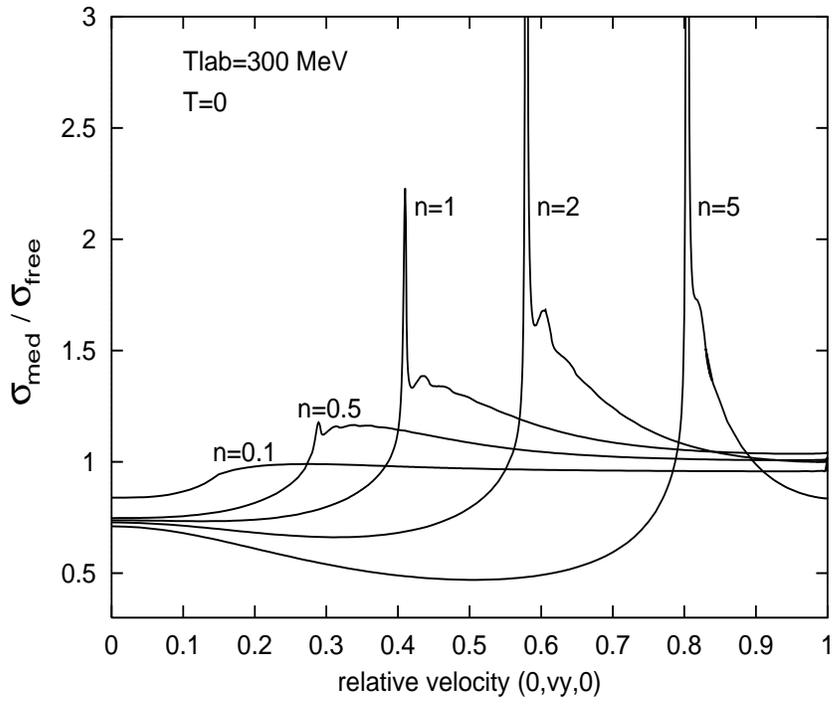
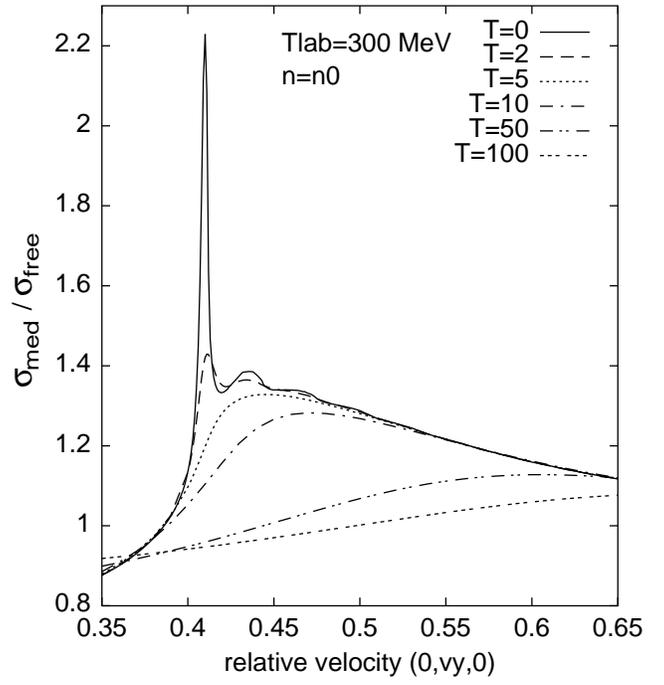
\end{document}